\documentclass{elsart}
\usepackage{graphicx,amssymb}
\journal{Physics Letters A}

\newcommand{\beq}{\begin{equation}}
\newcommand{\eeq}{\end{equation}}
\newcommand{\ba}{\begin{array}}
\newcommand{\ea}{\end{array}}
\newcommand{\bea}{\begin{eqnarray}}
\newcommand{\eea}{\end{eqnarray}}
\begin{document}
\begin{frontmatter}

\title{High-probability quantum state transfer among nodes of an open XXZ spin
chain}
\author{E.B.Fel'dman},
\ead{efeldman@icp.ac.ru}
\author{A.I.~Zenchuk \corauthref{cor}
}
\corauth[cor]{Corresponding author.}
\ead{zenchuk@itp.ac.ru}

\address{
Institute of Problems of Chemical Physics, Russian Academy of Sciences, Chernogolovka, Moscow reg., 142432, Russia}

\begin{abstract}
This paper concerns the problem of the high probability  state transfer  among $2^s$ symmetrically placed  nodes of the $N$-nodes spins 1/2 chain with the
  ${XXZ}$ Hamiltonian. We consider examples with $(N,s)=(4,4)$, $(N,s)=(6,4)$ and $(N,s)=(8,8)$.
\end{abstract}

\end{frontmatter}


\section{Introduction}

This paper concerns the problem of the  high probability state transfer (HPST) \cite{KZ} between different nodes of  the spin 1/2 chain, which becomes a popular problem   due to the development of the quantum communication systems and quantum computing.
By "state transfer" we mean the following phenomenon \cite{Bose,CDEL}.
Consider the   chain of spins 1/2 with dipole-dipole interactions in the strong external magnetic field.  Let  all spins be directed along the external magnetic  field except the $i$th one whose initial state is arbitrary,  $\psi_{i}= \cos \theta |0\rangle + e^{i\phi} \sin \theta |1\rangle$, where $|0\rangle$  and $|1\rangle $ mean the spin directed along and opposite the external magnetic field respectively. Let the energy of the ground state (all spins are aligned along the magnetic field) be zero. If  the state of $j$th node becomes $\psi_{j}= \cos \theta |0\rangle + e^{i\phi} f_{ij} \sin \theta |1\rangle$ with $|f_{ij}| = 1$ at time moment  $t_0$ then we say that the state has been  transfered from the $i$th to the $j$th node with the phase shift $\phi_{ij}=\arg f_{ij}$. Since $|f_{ij}|=1$, all other spins are directed along the field at $t=t_0$, i.e. their states are $|0\rangle $.  Function $f_{ij}$ is the transition amplitude  of  an excited state $|1\rangle$ from the $i$th to the $j$th node.  
 Note that if all nodes of the chain have equal Larmor frequencies and we are interested in the state propagation between two nodes, say between  $s$th to $r$th nodes, then the shift $\phi_{sr}$ may be simply  removed by the proper choice of the constant magnetic field value  \cite{Bose},  so that $f_{ij}=1$, i.e. the state is perfectly transfered. 
 
 There exists a wide literature studying   the state  transfer along the spin chains in the strong external magnetic field. For instance, propagation of the spin waves in homogeneous chains  was considered in
  \cite{FBE}. It was demonstrated that the state may be perfectly transfered between two end nodes \cite{CDEL,ACDE} as well as between    two symmetrical inner nodes \cite{KS} in the inhomogeneous chain. State propagation along the alternating chains was studied in \cite{FR,KF}. It was shown in \cite{VGIZ,GKMT} that  the chains with week end bonds provide the state transfer from one  to another side of the chain.   End-to-end entanglement in both  alternating chains and  chains with weak end bonds has been studied in\cite{VGIZ}. Some aspects of the entanglement between remote nodes of the chain have been  studied in \cite{FMS}.

However, all these references consider  the state transfer
between two end nodes (or between two symmetrical nodes) which is required for the construction of the communication channels where the state must be transfered from one object to another. Meanwhile, the quantum computation requires such  systems which have HPSTs among many different nodes and, as a  consequence, may distribute information among  these nodes. Such systems may be candidates for the quantum register. 
Emphasise that, as it was indicated above, the excited state may be transfered from the initial $i$th node to some $j$th node  with proper  phase shift $\phi_{ij}$. However, we will show that all these shifts may be removed  in a simple way introducing the time dependent magnetic field, see Sec.\ref{Section:XYZ} (remember that the single phase shift  can be removed by a  constant magnetic field, like it was done in the case of the state propagation between two nodes \cite{Bose}). 
Thus, in general, the phase shifts $\phi_{ij}$ do not create serious  obstacles for the quantum communications. The only problem is the organization of the state transfers with big values of $|f_{ij}|$. 

A simple variant of such systems is suggested in our paper. Namely, we construct the chain of $N$ nodes which  has set ${\cal{L}}$ of ${\cal{N}}\le N$ 
nodes  $p_i$, $i=1,\dots,{\cal{N}}$,
\begin{eqnarray}
{\cal{L}}=\{p_i, \;\;i=1,\dots,{\cal{N}}\le N\},\;\;\;p_1\equiv 1,
\end{eqnarray}
 with the HPSTs between any two of them, i.e.   
if the unknown state $\phi_{n}=\cos\theta |0\rangle + e^{i \phi}\sin\theta |1\rangle$   is generated in any particular node $n$ from the set ${\cal{L}}$ (while the initial states of all other nodes are $|0\rangle$) then this state    may be detected with high probability in any other node $m$ from the set ${\cal{L}}$ after  appropriate  time intervals.

\begin{figure*}[!htb]
\noindent    \resizebox{140mm}{!}{\includegraphics[width=14cm,angle=0]
{
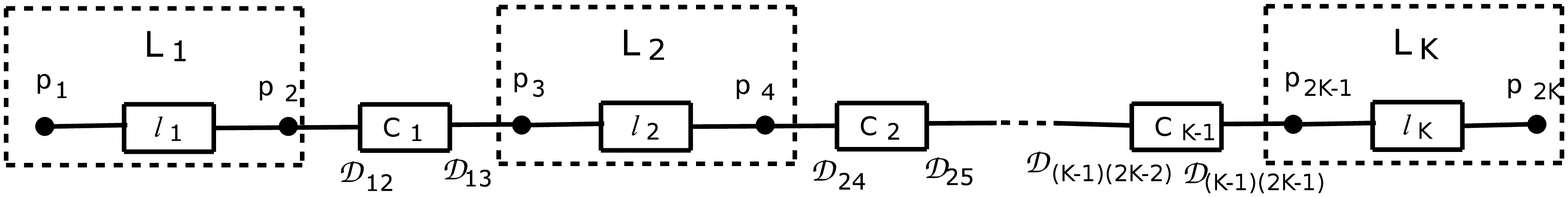}}
  \caption{The non-symmetric chain $L_{1,\dots,K}(N_1,M_1,N_2,M_2,\dots,N_K-1,M_K-1,N_K)$. The total number of the nodes is $N=\sum_{i=1}^{K-1}(N_i+M_i)+N_K$.   
  Here ${\cal{L}}=\{p_i,\;\;i=1,\dots,K\}$. The chains $L_{i}$ allow the HPSTs among their end nodes $p_{2i-1}$ and $p_{2i}$, $i=1,\dots,K$. These chains are connected through the spin 1/2  chains $C_{i}$, $i=1,\dots,K-1$. 
Parameters ${\cal{D}}_{i(2i)}$  (${\cal{D}}_{i(2i+1)}$)  are coupling constants between $p_{2i}$  ($p_{2i+1}$) and first (last) node of the chain $C_{i}$, $i=1,\dots,K-1$.  For unambiguity, the coupling constants between $n$th and $m$th nodes of the whole chain will be called $D_{nm}$. Namely parameters $D_{nm}$ appear in the Hamiltonian (\ref{Hamiltonian}). $D_n\equiv D_{n(n+1)}$ are coupling constants between the nearest neighbours.  }
  \label{Fig:gen}
\end{figure*}

Hereafter we will use the notation
$L_{1,\dots,K}(N_1,M_1,N_2,M_2,\dots,N_K-1,M_K-1,N_K)$ for the chain shown in Fig.\ref{Fig:gen}. 
Here $N_i$ and $M_i$ are  the numbers of nodes in $L_i$ and $C_i$
respectively. 
Let us clarify the structure of this scheme. Each of the chains $L_i$, $i=1,\dots,K$, allows the HPST between its end nodes $p_{2i-1}$ and $p_{2i}$. Chains $l_i$ collect all inner nodes of $L_i$ and consequently have $N_i-2$ nodes. Two chains $L_i$ and $L_{i+1}$ are connected by the "week bond"  through  the chain $C_i$, $i=1,\dots,K-1$.
By "week bond" between $L_i$ and $L_{i+1}$ we mean the following necessary  inequality  among the coupling constants:
\begin{eqnarray}\nonumber
&&
\max({\cal{D}}_{i(2i)},{\cal{D}}_{i(2i+1)}, {\mbox{coupling constants in $C_i$, $i=1,\dots,K-1$}}) < \\
&&
\min({\mbox{coupling constants in  $L_i$, $i=1,\dots,K$}}).
\end{eqnarray}

\begin{figure*}[!htb]
\noindent    \resizebox{140mm}{!}{\includegraphics[width=14cm,
angle=0]{
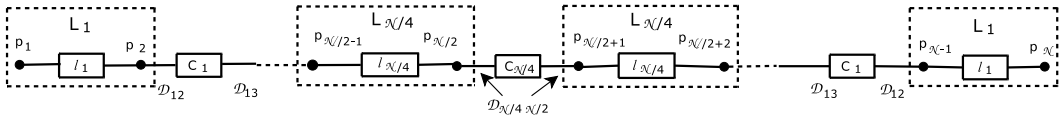}}
  \caption{ General scheme of the symmetric spin 1/2 chain with the HPSTs among the nodes $p_i$, $i=1,\dots,{\cal{N}}$, ${\cal{N}}=2^s$, $s=1,2,\dots$. Here the total number of the nodes  $N=2\sum_{i=1}^{{\cal{N}}/4-1}(N_i+M_i)+2 N_{{\cal{N}}/4}+M_{{\cal{N}}/4}$,   
  ${\cal{L}}=\{p_i,\;\;i=1,\dots,{\cal{N}}\}$.
  }
  \label{Fig:d1}
\end{figure*}

Emphasize that, also the HPST is organized  between end nodes of each particular chain $L_i$ (taken out of the general chain), the whole  chain $L_{1,\dots,K}$ does not provide the HPSTs between all $p_i$ ($i=1,\dots,2K$) in general case.  
However, we  are interested in the particular form of the chain $L_{1,\dots,K}$ which does  provide the HPST among all nodes $p_i$. 
First of all, such chain  
 must be symmetrical and may be written as  ($K={\cal{N}}/2$)
$$L_{1,\dots,{\cal{N}}/4-1,{\cal{N}}/4,{\cal{N}}/4,{\cal{N}}/4-1,
\dots,1}\Big(N_1,M_1,\dots, N_{\cal{N}}/4, M_{\cal{N}}/4, 
N_{\cal{N}}/4,\dots,M_1,N_1\Big),$$
 where ${\cal{N}}=2^s$, $s=1,2,\dots$, see Fig.\ref{Fig:d1}.

Let the chain have $N$ nodes. 
We use notations  $\bar P^{(N)}_{p_np_m}$, $\bar t^{(N)}_{p_n p_m}$ and $\bar \phi^{(N)}_{p_np_m}$ for the  probability of the exited state transfer between $p_n$th and $p_m$th nodes, for the  time interval  required for this transfer and for the phase shift of the transfered exited state, $n,m=1,\dots,{\cal{N}}\le N$:
\begin{eqnarray}
\bar P^{(N)}_{p_np_m}\equiv P^{(N)}_{p_np_m}(\bar t^{(N)}_{p_np_m})=|f_{p_ip_j}(\bar t^{(N)}_{p_np_m})|^2,\;\;\; \bar\phi^{(N)}_{p_ip_j}=\arg f^{(N)}_{p_ip_j}(\bar t^{(N)}_{p_np_m}).
\end{eqnarray}
Due to the symmetry of the chain, we have the following identities:
\begin{eqnarray}\label{symmetry}
&&
\bar P^{(N)}_{p_np_m}=\bar P^{(N)}_{p_mp_n},\;\;\bar t^{(N)}_{p_np_m}=\bar t^{(N)}_{p_mp_n},\;\; \bar\phi^{(N)}_{p_np_m}=\bar\phi^{(N)}_{p_mp_n}
\\\nonumber
&&
 \bar P^{(N)}_{p_np_m}=\bar P^{(N)}_{(N-p_n+1)(N-p_m+1)},\;\;
 \bar t^{(N)}_{p_np_m}=\bar t^{(N)}_{(N-p_n+1)(N-p_m+1)},\\\nonumber
 &&
\bar  \phi^{(N)}_{p_np_m}=\bar \phi^{(N)}_{(N-p_n+1)(N-p_m+1)}.
\end{eqnarray}

Because of the wide spread of the coupling constants, the time interval $\bar t^{(N)}_{p_np_m}$ needed for the state transfer between $p_n$ and $p_m$ significantly depends on the values $n$ and $m$. In general,
\begin{eqnarray}
\bar t^{(N)}_{p_np_m}|_{n\le {\cal{N}}/2,m>{\cal{N}}/2} \ll \bar t^{(N)}_{p_np_m}|_{n,m\le{\cal{N}}/2},
\end{eqnarray}
i.e. the state may be transfered between two nodes much faster if both nodes are placed in the same half of the chain. 
Thus, an important characteristic of such chain is the interval 
\begin{eqnarray}
 T^{(N)}_{{\cal{N}}}=
\max\Big(\bar t^{(N)}_{p_np_m},\;\;n,m=1,\dots {\cal{N}}\Big) .
\end{eqnarray}

Hereafter the state transfer between the nodes $p_n$ and $p_m$   will be referred to as HPST if
\begin{eqnarray}\label{HPST}
\bar P^{(N)}_{p_np_m}\ge P_0.
\end{eqnarray}
The value  $P_0$ is conventional. We take $P_0=0.9$ in  Examples of Sec.\ref{Section:4n} and in Example 1 of Sec.\ref{Section:def} and  $P_0=0.8$ in  Examples of Sec.\ref{Section:8n}
and in Example 2 of Sec.\ref{Section:def}.

The set of all possible HPSTs
 between any two nodes from the list ${\cal{L}}$ will be referred to as HPST$(N;{\cal{L}})$, where $N$ is the total number of nodes in the chain.
We call the parameters of HPST$(N;{\cal{L}})$ in such  chain  the set of parameters
\begin{eqnarray}\label{parametersHPST}
\bar P^{(N)}_{p_np_m}, \;\;\;\bar t^{(N)}_{p_np_m},\;\;\;\bar \phi^{(N)}_{p_np_m}.
\end{eqnarray}

Since the  organization of the HPSTs among different nodes of  the set of ${\cal{N}}>2$ 
nodes is an 
essential property of the quantum register, the chains 
constructed in this paper may be candidates for this role. 

While the nodes $p_i$ may serve as the q-bits of the quantum 
register, the chains    $l_i$ (and $C_i$) serve to decrease the time intervals $\bar  t^{(N)}_{p_{2i-1}p_{2i}}$ (and $\bar t^{(N)}_{p_{2i}p_{2i+1}}$)
 required for the state transfer between the nodes $p_{2i-1}$ and $p_{2i}$ (and between the nodes $p_{2i}$ and $p_{2i+1}$)
separated by the  long distance as it happens in the 
 communication channels. Namely, if $L_i$ consists of two nodes 
 $p_{2i-1}$ and $p_{2i}$ then the time interval  $\bar 
 t^{(N)}_{p_{2i-1}p_{2i}}$ may be reduced putting additional chain $l_i$ 
 with properly adjusted coupling constants between these two 
 nodes.  Similarly, the time interval $\bar t^{(N)}_{p_{2i}p_{2i+1}}$ 
 required to  transfer the excited state between the last node of
  $L_i$ (i.e. node $p_{2i}$) and the first node of $L_{i+1}$ 
  (i.e. node $p_{2i+1}$)  may be reduced putting chain $C_i$ with
  properly adjusted coupling constants between $L_i$ and 
  $L_{i+1}$, see, for instance, chain $C_1(2)$ in 
  Fig.\ref{Fig:d6_4n_mod_ex}. 
Thus the above spin chain $L_{1,\dots,{\cal{N}}/4-1,{\cal{N}}/4,{\cal{N}}/4,{\cal{N}}/4-1,
\dots,1}\Big(N_1,M_1,\dots, N_{\cal{N}}/4, M_{\cal{N}}/4, 
N_{\cal{N}}/4,\dots,M_1,N_1\Big)$ combines properties of both quantum register
and communication channel.  

Finally we note that,
constructing the spin chain,  we want 
\begin{eqnarray}\label{condition1}
(1)&&{\mbox{
to satisfy the condition (\ref{HPST}) for  $\bar P^{(N)}_{p_np_m}$, $n,m=1,\dots,{\cal{N}}$}},\\\label{condition2}
(2)&&
{\mbox{to minimize the parameter $T^{(N)}_{{\cal{N}}}$}},
\\\label{consition3}
(3)&&
{\mbox{to remove phase shifts $\bar\phi^{(N)}_{p_np_m}$, $n,m=1,\dots,{\cal{N}}$}}.
\end{eqnarray}

This paper is organized as follows.
In Sec.\ref{Section:XYZ} we show that $2^N\times 2^N$ matrix representation of the   XXZ 
Hamiltonian  may be reduced to  the $N\times N$ matrix representation  for a
 certain type of problems. Sec.\ref{Section:4n} describes the
  general structures of the spin chains having the set ${\cal{L}}$ of four 
  nodes;  simple examples are represented. Similar study of the eight 
  node chain is given in Sec.\ref{Section:8n}. Deformations of 
  the above chains decreasing parameter $T^{(N)}_{{\cal{N}}}$ are discussed
  in Sec.\ref{Section:def}.  Conclusions are given in 
  Sec.\ref{Section:Conclusions}.

\section{State transfer in spin 1/2 chain with $XXZ$ Hamiltonian}
\label{Section:XYZ}
We study the HPSTs  \cite{KZ} among   nodes of the spin 1/2 chain in strong external magnetic field $B(t)$ with the $XXZ$
 Hamiltonian
\begin{eqnarray}\label{Hamiltonian}
{\cal{H}}&=&{\cal{H}}_{dz}+ \omega(t) I_z,\;\;\;I_z=\sum_{i=1}^N I_{i,z},\;\;\;\omega(t)= \gamma B(t)\\\nonumber
{\cal{H}}_{dz} &=&
\sum_{{i,j=1}\atop{j>i}}^{N}
D_{ij}(I_{i,x}I_{j,x} + I_{i,y}I_{j,y}-2  I_{i,z}I_{j,z}),\;\;\;
D_{ij}=\frac{\gamma^2 \hbar}{r_{ij}^3},
\end{eqnarray}
where $\gamma$ is the gyromagnetic ratio,  $r_{ij}$ is the distance between $i$th and $j$th spins, $I_{i,\alpha}$ is the projection operator of the $i$th  spin on the $\alpha$ axis, $\alpha=x,y,z$, $I_z$ is the   $z$-projection operator of the total spin,
$D_{ij}$  are the spin-spin coupling constants.
This Hamiltonian describes  the secular part of the dipole-dipole interaction in the strong external  magnetic field
\cite{A}.
For our convenience, we call  $D_i\equiv D_{i(i+1)}$, $i=1,\dots,N-1$. It is obvious that all $D_i$ may be arbitrary by definition.

\begin{figure*}[!htb]
\noindent    \resizebox{140mm}{!}{\includegraphics[width=14cm,angle=0]
{
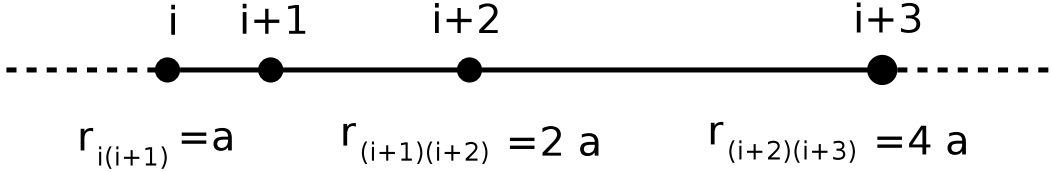}}
  \caption{Example of the chain which does not allow the approximation by the nearest neighbour interaction}
  \label{Fig:chain}
\end{figure*}

Although the approximation of the Hamiltonian (\ref{Hamiltonian}) by the nearest neighbour interaction  is very popular in  study of the quantum state transfer, one can show that it is not  applicable to the inhomogeneous chains which have a wide spread of  coupling constants.
In fact, this approximation is applicable if
$D_{ij}\ll \min(D_{k},\;k=1,\dots,N-1)$, $\forall$  $j>i+1$. Only in this case we may disregard terms with coupling constants $D_{ij}$ ($j>i+1$) in the Hamiltonian. However, this is not possible in general.
Consider, for instance, the chain in Fig.\ref{Fig:chain}.
In this chain, the approximation by the nearest neighbour interaction takes into account the term with coupling constant $D_{i+2}=\frac{\alpha}{64 a^3}$ and disregards the term with coupling constant $D_{i(i+2)}=\frac{\alpha}{27 a^3}>D_{i+2}$ in the Hamiltonian, i.e. we neglect the term which is bigger then the term  which is taken into account.
Comparison of the coupling constants in the chains 
considered in  Secs.\ref{Section:4n}-\ref{Section:def} confirms that the 
approximation of the Hamiltonian by the nearest neighbour 
interaction is not applicable to these chains. Thus, hereafter we
consider the total Hamiltonian (\ref{Hamiltonian}).
 
 It is convenient to take the eigenvectors of $I_z$  as  the basis of the matrix representation of the Hamiltonian (\ref{Hamiltonian}). This is possible since  the Hamiltonian (\ref{Hamiltonian}) commutes with $I_z$:
\begin{eqnarray}
[{\cal{H}},I_z]=0,
\end{eqnarray}
so that both ${\cal{H}}$ and $I_z$ have the common set of eigenvectors.
In general the  dimensionality of the matrix representation of the Hamiltonian is $2^N\times 2^N$.
 As usual, we  write the eigenvectors of the operator $I_z$ in terms of the Dirac notations.  Let
\begin{eqnarray}
|n_1\dots n_N \rangle
\end{eqnarray}
be the eigenvector of the operator $I_z$ where  the $i$th spin is directed opposite to the external magnetic field if $n_i=1$ and  along the field if $n_i=0$.
Then the matrix representation $H$ of the Hamiltonian ${\cal{H}}$  gets the following diagonal block structure:
\begin{eqnarray}\label{H_diag}
{{H}}={\mbox{diag}}(H_0,H_1,H_2,\dots,H_N),
\end{eqnarray}
where the block $H_i$ is assotiated with
the states having $i$ spins directed opposite to the field. The dimensionality if this  block is $C_N^i\times C_N^i$.

It is important that in order to study  the problem of the single quantum state transfer along the spin 1/2 chain with Hamiltonian (\ref{Hamiltonian}) only the blocks $H_0$ and $H_1$ are needed,
  \begin{eqnarray}\label{H1}
&&H_0=-\frac{1}{2}\Big(\tilde\gamma- N \omega(t)\Big) ,\\\nonumber
&&
H_1=\frac{1}{2}\Big[D -\Big(\tilde\gamma-(N-2)\omega(t)\Big) I\Big],\\\nonumber
&&
D=\left(\begin{array}{ccccccc}
A_{11} & D_1 &D_{13} & \cdots& D_{1(N-2)} & D_{1(N-1)}&D_{1N}\cr
D_1& A_{22} & D_2 & \cdots& D_{2(N-2)} & D_{2(N-1)}&D_{2N}\cr
D_{13}&D_2& A_{33} & \cdots& D_{3(N-3)} & D_{3(N-1)}&D_{3N}\cr
\vdots &\vdots &\vdots &\vdots&\vdots &\vdots &\vdots \cr
D_{1(N-2)}&D_{2(N-2)}&D_{3(N-2)}&\cdots&A_{(N-2)(N-2)}&D_{j-1}&D_{(N-2)N}\cr
D_{1(N-1)}&D_{2(N-1)}&D_{3(N-1)}&\cdots&D_{j-1}&A_{(N-1)(N-1)}&D_j\cr
D_{1N}&D_{2N}&D_{3N}&\cdots&D_{(N-2)N}&D_j&A_{NN}
\end{array}\right),
\\\nonumber
&&
A_{nn}=2\sum_{{i=1}\atop{i\neq n}}^{N}D_{in},\;\;\tilde\gamma=\sum_{{i,j=1}\atop{i< j}}^{N}D_{ij},
\end{eqnarray}
where $I$ is $N\times N$ identity matrix.
Thus, instead of $2^N\times 2^N$ matrix representation of the Hamiltonian we have $N\times N$ block $H_1$ and scalar block $H_0$, which significantly reduces all calculations.

It was shown \cite{Bose} that effectiveness of the quantum communication channel may be measured by the fidelities of the state transfers between $p_n$th and $p_m$th nodes:
\begin{eqnarray}\label{F_nm}
F^{(N)}_{p_np_m}(t)=\frac{|f^{(N)}_{p_np_m}(t)| \cos \Gamma^{(N)}_{p_np_m}(t) }{3} + \frac{|f^{(N)}_{p_np_m}(t)|^2}{6} +\frac{1}{2},
\end{eqnarray}  
where the amplitudes $f^{(N)}_{p_np_m}$ and the phases $\Gamma^{(N)}_{p_np_m}$ are defined as follows ($n,m=1,\dots,{\cal{N}}$):
\begin{eqnarray}\label{f_nm}
&&
f^{(N)}_{p_np_m}(t)=\langle p_m |e^{-i{\cal{H}} t}|p_n\rangle 
=\sum_{j=1}^{N}  
u_{p_nj} u_{p_mj}e^{-i\lambda_j  t/2} e^{\frac{i}{2}\left(\tilde\gamma t- (N-2)\int\limits_{0}^t \omega(\tau) d \tau\right)}
,\\\label{Gamma}
&&
\Gamma^{(N)}_{p_np_m}(t)=\varphi^{(N)}_{p_np_m}(t)-\int\limits_0^t \omega(\tau)d\tau ,\\\label{varphi}
&&
 \varphi^{(N)}_{p_np_m}(t)=\arg\left(\sum_{j=1}^{N}  
u_{p_nj} u_{p_mj}e^{-i\lambda_j  t/2}\right).
\end{eqnarray}
Here $u_{ij}$, $i,j,=1,\dots,N$, are components of the eigenvector
$u_j$ corresponding to the eigenvalue $\lambda_j$ of the matrix $D$:
$Du_j=\lambda_j u_j$. The effectiveness of the state transfer is characterized by the set of parameters $\bar F^{(N)}_{p_np_m}\equiv F^{(N)}_{p_np_m}(t^{(N)}_{p_np_m})$, $n,m=1,\dots,{\cal{N}}$. It is evident, that these parameters take maximum values if
\begin{eqnarray}\label{Gamma0}
\bar\Gamma^{(N)}_{p_np_m}\equiv \Gamma^{(N)}_{p_np_m}(\bar t^{(N)}_{p_np_m}) =0 \;\;\;{\mbox{mod}} (2\pi),\;\;\;n,m=1,\dots,{\cal{N}}.
\end{eqnarray}
which may be considered as the system of  equation defining (not uniquely) the time dependence  $\omega(t)$ (which must be positive over the interval $0\le t\le T^{(N)}_{\cal{N}}$):
\begin{eqnarray}\label{Gamma_B}
&&
\bar \omega^{(N)}_{p_np_m}\equiv \int\limits_0^{\bar t^{(N)}_{p_np_m}} \omega(\tau) d\tau=\bar\varphi^{(N)}_{p_np_m}\equiv \varphi^{(N)}_{p_np_m}(\bar t^{(N)}_{p_np_m}) \;\;\;{\mbox{mod}}(2\pi),\\\nonumber
&&
n,m=1,\dots,{\cal{N}}.
\end{eqnarray}
Then the values of  the fidelities $F^{(N)}_{p_np_m}(\bar t^{(N)}_{p_np_m})$ are defined by $|f^{(N)}_{p_np_m}(\bar t^{(N)}_{p_np_m})|$ or, equivalently,  by the probabilities $\bar P^{(N)}_{p_np_m}$. For this reason, namely probabilities $\bar P^{(N)}_{p_np_m}$ will be studied in the rest of this paper.
Phases $\varphi^{(N)}_{p_np_m}(t)$ (which are independent on  $\omega(t)$) will be taken as parameters of the HPST$(N;{\cal{L}})$  instead of the parameters $\bar\phi^{(N)}_{p_np_m}$  in the set (\ref{parametersHPST}), see  Tables 1 - 6.
Remark that, deriving eqs.(\ref{F_nm}-\ref{varphi}) we took into account that the energy of the ground state $E_0$ is not zero: $E_0=\hbar H_0$.

An explicite example of the  function $\omega(t)$ constructed in accordance with eqs.(\ref{Gamma_B})  with $N={\cal{N}}=4$ will be given in Sec.\ref{Section:4n}, see Example 1.

\section{Simplest chains allowing the HPSTs among the four nodes.}
\label{Section:ch1}
\label{Section:4n}
\begin{figure*}[!htb]
\noindent    \resizebox{350mm}{!}{\includegraphics[width=35cm,
angle=0]{
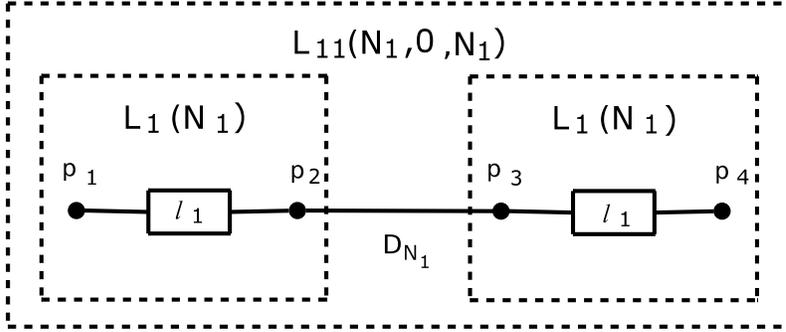}}
  \caption{ Spin 1/2 chain $L_{11}(N_1,0,N_1)$ with the HPSTs among the four nodes $p_i$, $i=1,2,3,4$, i.e.  ${\cal{L}}=\{p_i,\;\;i=1,2,3,4\}$, ${\cal{N}}=4$;  $C_1$ is absent.
  }
  \label{Fig:d2_4n}
\end{figure*}

The HPSTs from the end node to (some of) the inner nodes of the spin chain
becomes possible due to the following mechanism. Let us take two
identical chains $L_1(N_1)$  which allow the HPST between their end
nodes.  Let $\bar D$ be the  minimal coupling constant between the
nearest neighbours in these chains. We connect these chains by
the weak bond with the coupling constant $D_{N_1}\ll\bar D$,
resulting in  the chain $L_{11}(N_1,0,N_1)$ of $N=2 N_1$ nodes, see Fig.\ref{Fig:d2_4n}.
Then 
\begin{eqnarray}
{\cal{L}}=\{{\mbox{end nodes of chains $L_1(N_1)$ }}\} =
\{p_1,p_2,p_3,p_4\},\;\;{\cal{N}}=4 .
\end{eqnarray}
These HPSTs may be understood as follows. Let spin $p_1$   be directed opposite to the external magnetic field  while all other spins of the chain $L_{11}(N_1,0,N_1)$ be directed along the field initially. Due to the small coupling constant  $D_{N_1}$ this excited state remains inside of the first chain  $L_1{(N_1)}$ for the long time with the high probability to be detected in either $p_1$ or $p_2$  (since, by definition,  the chain $L_1{(N_1)}$ provides HPST between end nodes). However, due to the bond between two chains $L_1{(N_1)}$, the excited state  will  be transfered  to the second chain $L_1(N_1)$ after comparatively long time interval $T_{tr}$ with the high probability. If this happens, than the excited state remains in this chain for the long time with  high probability to be detected in either $p_3$ or $p_4$. Similarly, after one more time interval $T_{tr}$ the excited state will return to the first chain $L_1(N_1)$ with the high probability,  and so on. Parameter $D_{N_1}$ may be fixed by the two conditions (\ref{condition1}) and (\ref{condition2}).
 Hereafter in this section we take $P_0=0.9$ in the definition (\ref{HPST}). Consider two simple examples.

\begin{figure*}[!htb]
\noindent    \resizebox{100mm}{!}{\includegraphics[width=10cm,
angle=0]{
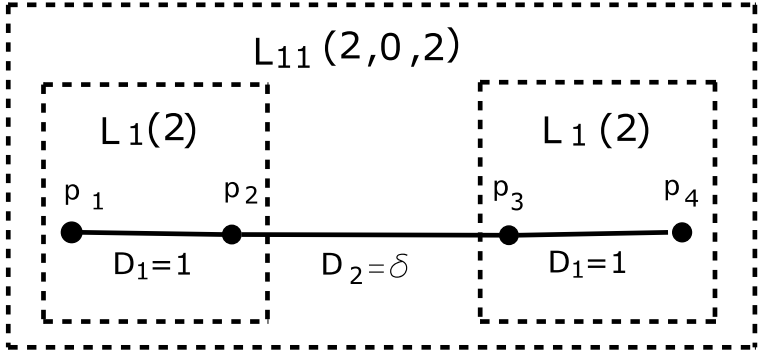}}
  \caption{ Spin 1/2 chain $L_{11}(2,0,2)$ with the HPSTs among all nodes $p_i\equiv i$, $i=1,2,3,4$, i.e.  ${\cal{L}}=\{1,2,3,4\}$,
  {\cal{N}}=4. This figure is equivalent to Fig.\ref{Fig:d2_4n} with $N_1=2$. The optimal value of the parameter $\delta$ is $0.196$. 
  }
  \label{Fig:d3_4n_ex1}
\end{figure*}

{\bf Example 1: the HPSTs in the chain $L_{11}(2,0,2)$.}
The simplest example of the chain $L_1{(N_1)}$ providing the ideal end-to-end
 state transfer is the chain of two nodes, i.e. $N_1=2$.
Connecting two equivalent  chains $L_1(N_1)$ we obtain  the chain  $L_{11}(2,0,2)$   with $N={\cal{N}}=4$, see Fig.\ref{Fig:d3_4n_ex1}. 

\begin{figure*}[!htb]
\noindent    
\resizebox{70mm}{!}{\includegraphics[width=5cm,angle=270]
{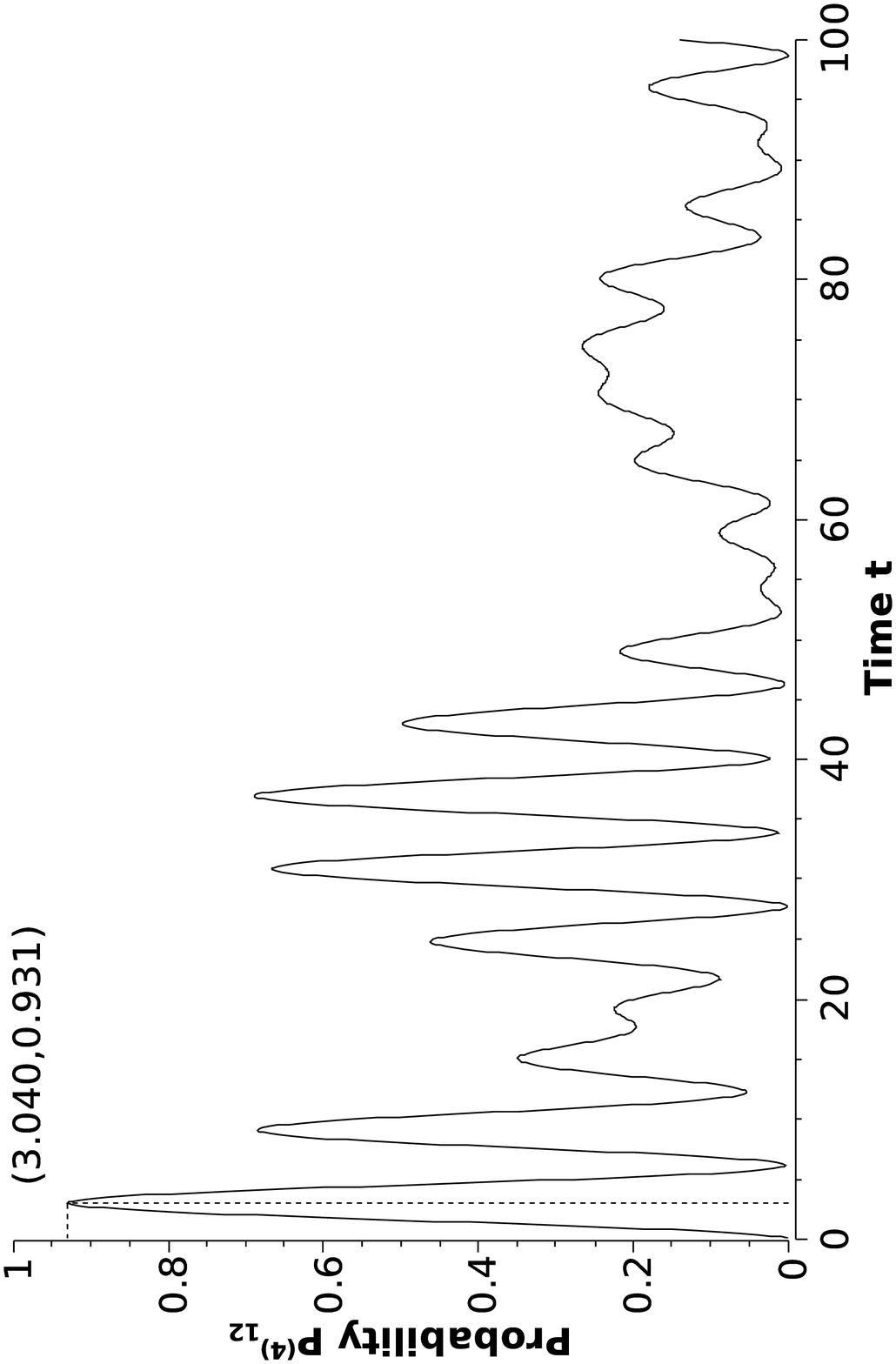}}
    \hfill 
\resizebox{70mm}{!}{\includegraphics[width=5cm,angle=270]
{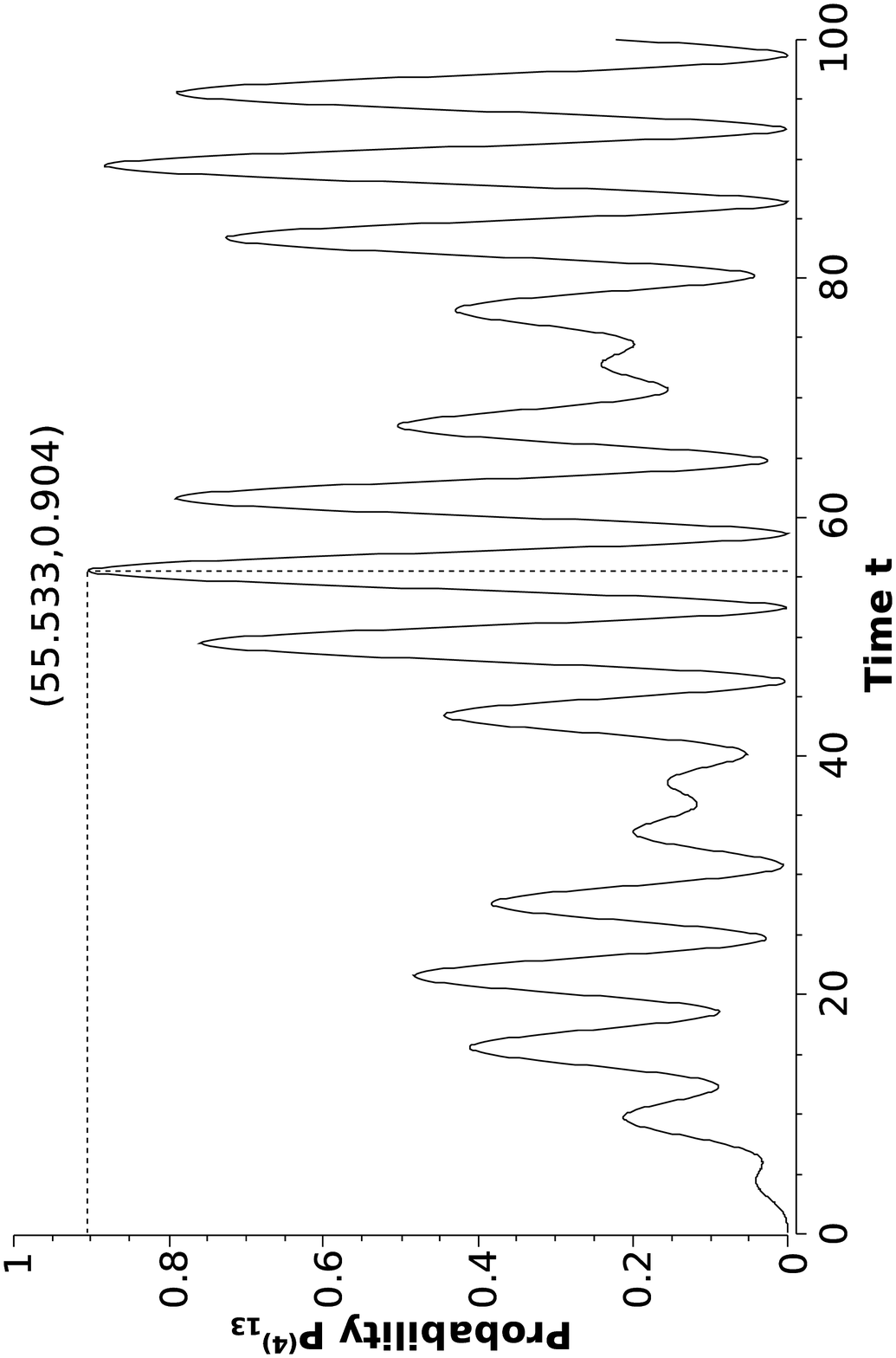}}
   \hfill  
   \resizebox{70mm}{!}{\includegraphics[width=5cm,angle=270]
   {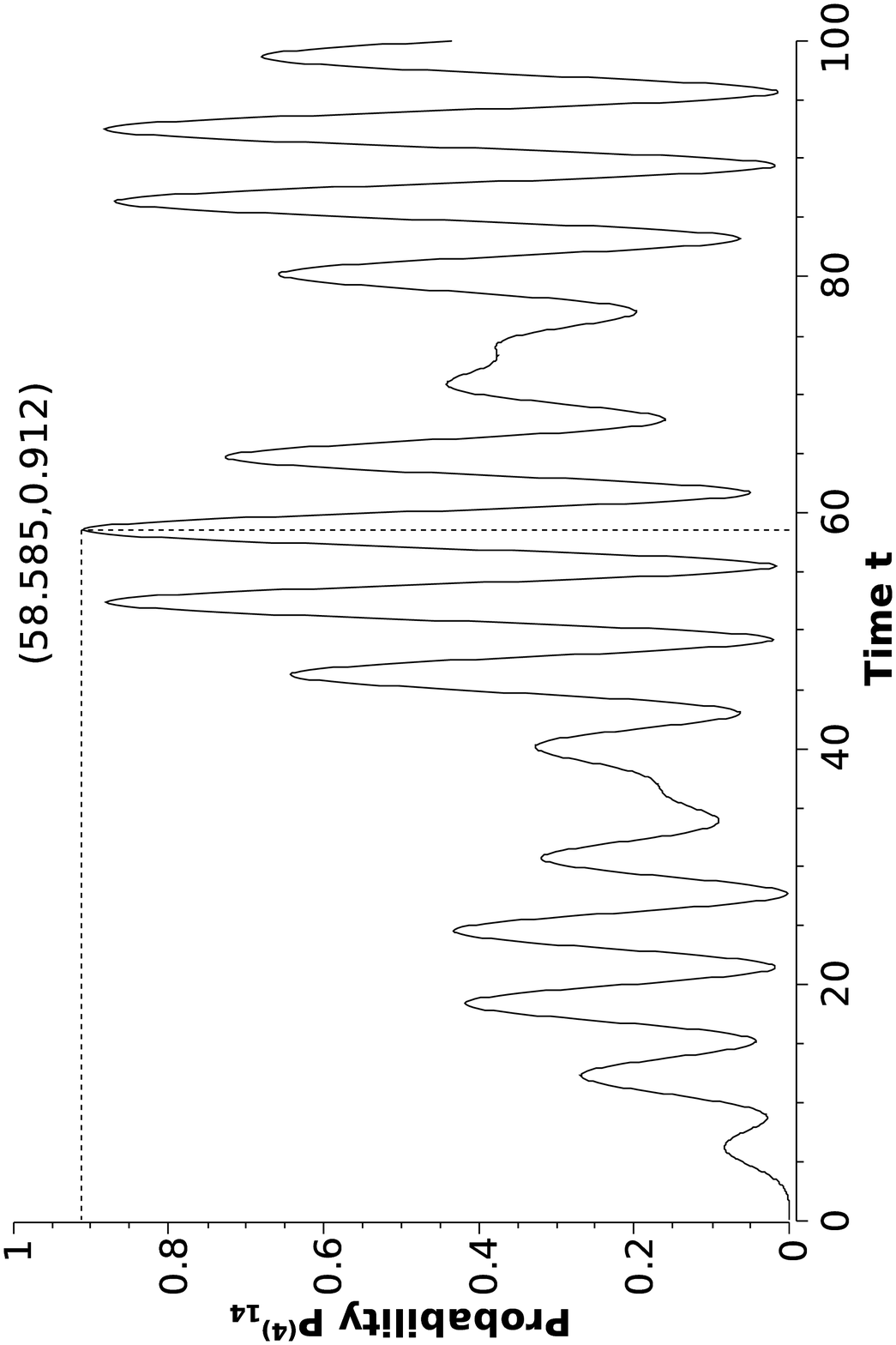}}
   \hfill 
   \resizebox{70mm}{!}{\includegraphics[width=5cm,angle=270]
   {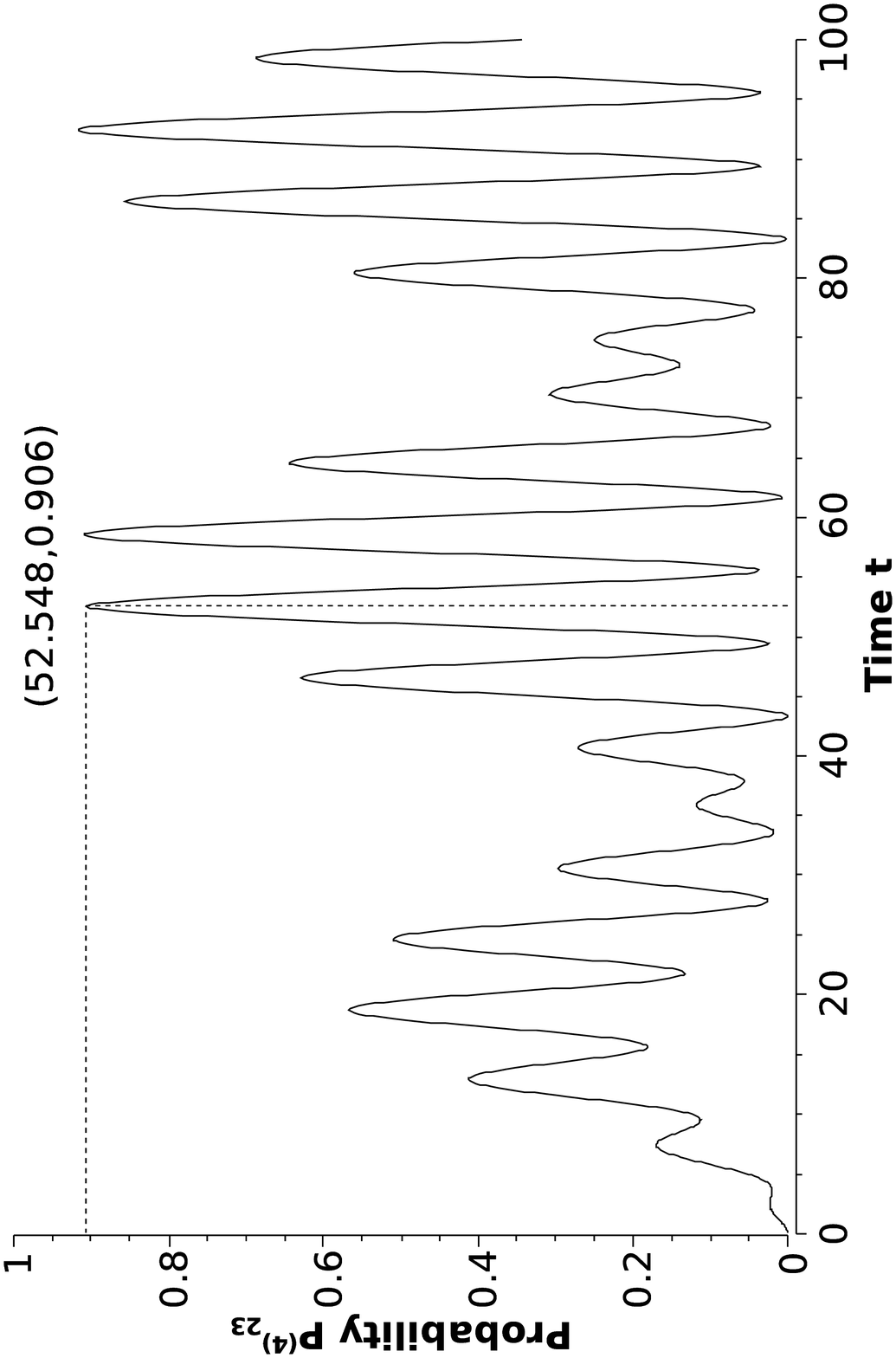}}
   \hfill
  \caption{The time dependence of the probabilities $P^{(4)}_{1i}$, $i=1,2,3,4$ and $P^{(4)}_{23}$ in the chain 
   $L_{11}(2,0,2)$.  The marked points correspond to the parameters $\bar P^{(4)}_{ij}$ and $\bar t^{(4)}_{ij}$ of the HPST(4;1,2,3,4), see also Table 1.
  }
  \label{Fig:N4}
\end{figure*}

In this case the set ${\cal{L}}$ consists of all nodes of the chain $L_{11}(2,0,2)$.
We  set $D_1=D_3=1$ for simplicity and vary $D_2=\delta$ in order to satisfy the conditions (\ref{condition1}) and (\ref{condition2}).
We have found that the optimal  parameters of the HPST(4;1,2,3,4) correspond to $\delta=0.196$. These parameters are represented in  Table 1, see also Fig.\ref{Fig:N4}. 

Let us find the time evolution of the external magnetic field which satisfies conditions (\ref{Gamma_B}), i.e such $\omega(t)$ that 
\begin{eqnarray}\label{Gamma0_2}
\bar \omega^{(4)}_{nm}\equiv \int\limits_0^{\bar t^{(4)}_{nm}} \omega(\tau) d\tau=\bar\varphi^{(4)}_{nm}\;\;\;{\mbox{mod}}(2\pi),\;\;\;n,m=1,2,3,4.
\end{eqnarray}  
We may  write conditions (\ref{Gamma0_2}) as the following system of five equations:
\begin{eqnarray}
&&
\bar \omega^{(4)}_{12}=0,\;\;\;
\bar \omega^{(4)}_{12}=\bar \varphi^{(4)}_{12},\;\;\;
\bar \omega^{(4)}_{13}=\bar \varphi^{(4)}_{13}+8 \pi,\\\nonumber
&&
\bar \omega^{(4)}_{14}=\bar \varphi^{(4)}_{14}+8 \pi,\;\;\;
\bar \omega^{(4)}_{23}=\bar \varphi^{(4)}_{23}+6 \pi,
\end{eqnarray}
which is satisfied by the   following function $\omega(t)$:
\begin{eqnarray}
\int\limits_0^t \omega(\tau)d\tau &=&a_4 t^4 +a_3 t^3+a_2 t^2 +a_1 t\;\;\;\Rightarrow \\\nonumber
\omega(t) &=&4 a_4 t^3 +3 a_3 t^2+2 a_2 t +a_1,\\\nonumber
&&
a_1=4.1241 \times 10^{-1},\;\;\;a_2=-6.9297 \times 10^{-3},\\\nonumber
&&
a_3=2.3114 \times 10^{-4},\;\;a_4=-1.9971 \times 10^{-6}.
\end{eqnarray}
We see that the function $\omega(t)$ constructed in this way is positive over the interval $0\le t \le \bar t^{(4)}_{14}=\max\limits_{n,m=1,\dots,4}(\bar t^{(4)}_{nm})$.

\begin{figure*}[!htb]
\noindent    \resizebox{140mm}{!}{\includegraphics[width=14cm,
angle=0]{
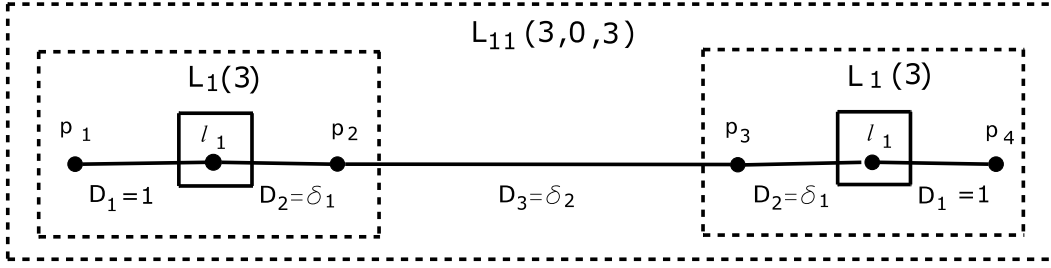}}
  \caption{ Spin 1/2 chain $L_{11}(3,0,3)$ with the HPSTs among  the four nodes $p_1=1$, $p_2=3$, $p_3=4$, $p_4=6$; ${\cal{L}}=\{1,3,4,6\}$, ${\cal{N}}=4$. $\delta_1=1$, $\delta_2=0.028$ in  Example 2 of Sec.\ref{Section:4n} and $\delta_1=0.769$, $\delta_2=0.092$ in the chain $\hat L_{11}(3,0,3)$ considered in Example 1 of Sec.\ref{Section:def}
  }
  \label{Fig:d3_4n_ex2}
\end{figure*}

{\bf Example 2: the HPSTs in the chain $L_{11}(3,0,3)$.}
We take two homogeneous chains of three nodes
$L_1(3)$. It is known that the ideal state transfer is
possible between the end nodes of these chains \cite{CDEL}.
We  connect them  by the weak bond obtaining the chain
shown in Fig.\ref{Fig:d3_4n_ex2}. 
Thus, ${\cal{L}}=1,3,4,6$.
Let $D_1=D_2=D_4=D_5=1$, $D_3=\delta_2$ in this chain.
We  vary $\delta_2$ to obtain the best correspondence to the
conditions (\ref{condition1}) and (\ref{condition2}).
The optimal value is $\delta_2=0.028$. The appropriate parameters of the
HPST(6;1,3,4,6) are represented in 
Table 2.

  
\subsection{Modification of the chain $L_{11}(N_1,0,N_1)$}
\label{Section:ch2}
\label{Section:4n_mod}
The chain shown in Fig.\ref{Fig:d2_4n} is convenient for the state transfer between two chains  $L_1$  if only the  distance between them  (i.e. between $p_2$ and $p_3$) is not too long. Otherwise the time interval $T^{(2N_1)}_{4}$  becomes very long. To decrease  this time interval we suggest the following modification of the chain $L_{11}(N_1,0,N_1)$.

\begin{figure*}[!htb]
\noindent    \resizebox{300mm}{!}{\includegraphics[width=30cm,angle=0
]{
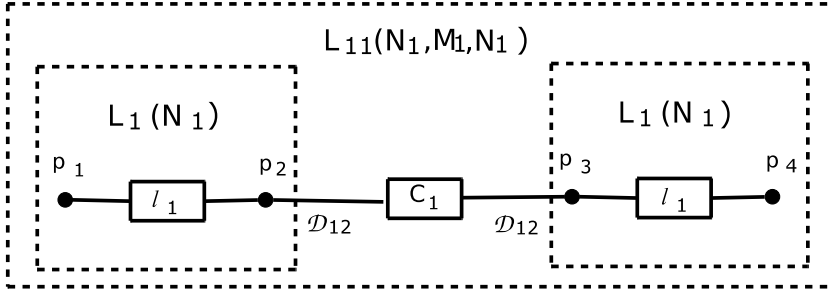}}
  \caption{ Spin 1/2 chain $L_{11}(N_1,M_1,N_1)$ with the HPSTs among  the four nodes $p_i$, $i=1,2,3,4$;  ${\cal{L}}=\{p_1,p_2,p_3,p_4\}$, ${\cal{N}}=4$. Chain $C_1(M_1)$   is introduced to decrease the parameter $T^{(N)}_{4}$, $N=2 N_1+M_1$ (compare with the  scheme in Fig.\ref{Fig:d2_4n}). 
  }
  \label{Fig:d5_4n_mod}
\end{figure*}

Let us take a  symmetrical  chain $C_1{(M_1)}$  of $M_1$ nodes with maximal coupling constant between neighbours  ${\cal{D}}$ satisfying the following condition: ${\cal{D}}< \bar D$. Using the coupling constant  
${\cal{D}}_{12}\ll \bar D$ we may construct the chain $L_{11}(N_1,M_1,N_1)$ of $N=2 N_1 +M_1$ nodes shown in Fig.\ref{Fig:d5_4n_mod}.
Here ${\cal{L}}$ consists of the end nodes of chains $L_1$ and may not involve any node of $C_1(M_1)$. This statement is valid   due to the fact, that the probability for the spin to be detected in the chain  $C_1{(M_1)}$ may not be high because of the small coupling constants both inside of this chain and 
${\cal{D}}_{12}$.
These coupling constants should be fixed by the conditions (\ref{condition1}) and (\ref{condition2}).  

\begin{figure*}[!htb]
\noindent    \resizebox{140mm}{!}{\includegraphics[width=14cm,angle=0]{
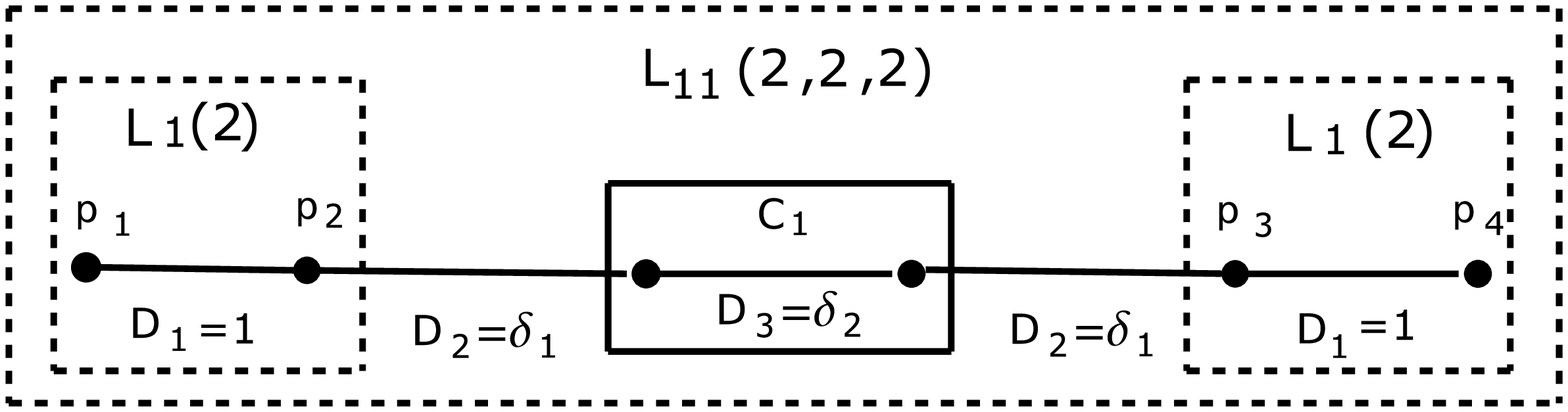}}
  \caption{ Spin 1/2 chain $L_{11}(2,2,2)$ with the HPSTs among  the four nodes $p_1=1$, $p_2=2$, $p_3=5$, $p_4=6$; ${\cal{L}}=\{1,2,5,6\}$, ${\cal{N}}=4$. Chain $C_1(2)$ is introduced to decrease the parameter $T^{(6)}_{4}$ (compare with  the scheme in  Fig.\ref{Fig:d5_4n_mod}). 
 Optimal parameters of the HPST(6;1,2,5,6) correspond to $\delta_1=0.224$, $\delta_2=0.649$.
  }
  \label{Fig:d6_4n_mod_ex}
\end{figure*}

{\bf Example: the HPSTs in the chain $L_{11}(2,2,2)$.}
We consider the  HPST(6;1,2,5,6) in the chain
$L_{11}(2,2,2)$ shown in Fig.\ref{Fig:d6_4n_mod_ex}.  

This is a chain of 6 nodes
 ($N=6$). We fix $D_1=D_5=1$ and vary parameters
 $D_2=D_4=\delta_1$ and $D_3=\delta_2$ with the purpose to 
decrease the  parameter $T^{(6)}_{4}$ and  satisfy the conditions (\ref{condition1},\ref{condition2}).
 We have found that the following values of the parameters $\delta_i$  yield a good result:
 $\delta_1=0.224$ and $\delta_2=0.649$.
 The appropriate parameters of the HPST(6;1,2,5,6) are represented in  Table 3.

Here we demonstrate that the  intermediate chain
 $C_1(2)$ in the chain $L_{11}(2,2,2)$  speeds up the
 state transfer between $p_2$ and $p_3$ separated by the distance
$R=2 (\Gamma/\delta_1)^{1/3}+(\Gamma/\delta_2)^{1/3}$.
For this purpose we compare the parameters $\bar t^{(4)}_{23}$ 
for the chain $L_{11}(2,0,2)$ (see Fig.\ref{Fig:d3_4n_ex1}) with 
$D_2=\Gamma/R^3\approx 0.011$ and parameter $\bar t^{(6)}_{25}\approx 64.4$ 
from the Table 3. 
Numerical simulation shows that  $\bar t^{(4)}_{23}\approx  2000$, i.e.
 $\bar t^{(4)}_{23}/\bar t^{(6)}_{25}\approx 31$.

\section{Simplest chains with the HPSTs among eight nodes.}
\label{Section:ch3}
\label{Section:8n}
Chains $L_{11}(N_1,0,N_1)$ and $L_{11}(N_1,M_1,N_1)$ considered in Sec.\ref{Section:4n}  provide the HPSTs between any two nodes out of the set ${\cal{L}}$ consisting of four nodes. However, the number of such nodes may be
 increased using the following obvious generalization of these
 chains.  

\begin{figure*}[!htb]
\noindent    \resizebox{140mm}{!}{\includegraphics[width=14cm,
angle=0]{
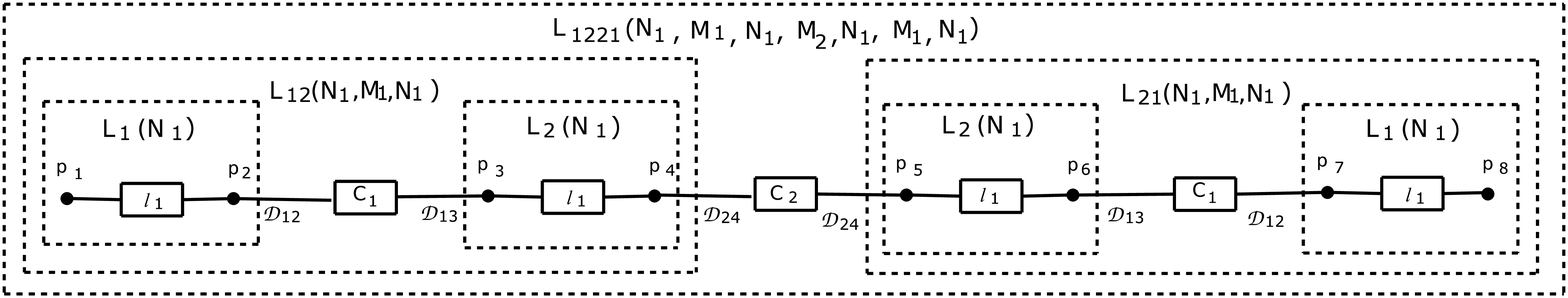}}
  \caption{ Spin 1/2 chain $L_{1221}(N_1,M_1,N_1,M_2,N_1,M_1,N_1)$ with the HPSTs among  the eight nodes $p_i$, $i=1,\dots,8$;  ${\cal{L}}=\{p_i,\;\;i=1,\dots,8\}$, ${\cal{N}}=8$. 
  }
  \label{Fig:8n}
\end{figure*}

Let us take two  chains  $L_{12}(N_1,M_1,N_1)$
and $L_{21}(N_1,M_1,N_1)$  (see Fig.\ref{Fig:gen}) and the symmetrical chain $C_2(M_2)$. The
maximal coupling constant $\bar{\cal{D}}$ between the nearest neighbours in $C_2(M_2)$ must satisfy inequality $\bar{\cal{D}}<\min( {\cal{D}}_{12}, {\cal{D}}_{13},\bar D)$, where $\bar D$ is the minimal coupling constant between the nearest neighbours in $L_{i}(N_1)$, $i=1,2$.  Using the coupling constant ${\cal{D}}_{24}$,
${\cal{D}}_{24}\ll \bar{\cal{D}}$,
we  construct the chain of $N=4 N_1 + 2 M_1 + M_2$ nodes shown in Fig.\ref{Fig:8n}.
Here ${\cal{L}}$ consists of the eight end nodes of chains $L_1$ and $L_2$: ${\cal{L}}=\{p_i,i=1,\dots,8\}$, ${\cal{N}}=8$.  

It is obvious that this algorithm may be extended to construct chains with ${\cal{L}}$ consisting of   $2^s$ nodes $s=1,2,\dots$..
Consider the simplest example where we take $P_0=0.8$ (see eq.(\ref{HPST})).

\begin{figure*}[!htb]
\noindent    \resizebox{140mm}{!}{\includegraphics[width=14cm,
angle=0]{
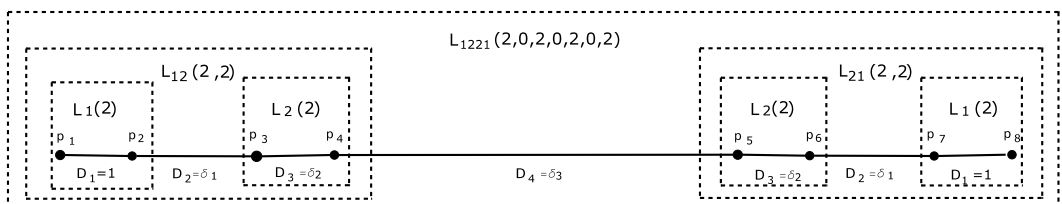}}
  \caption{ Spin 1/2 chain $L_{1221}(2,0,2,0,2,0,2)$ with the HPSTs among  the eight nodes $p_i=i$, $i=1,\dots,8$;  ${\cal{L}}=\{1,\dots,8\}$, ${\cal{N}}=8$; $\delta_1=0.196$, $\delta_2=1$, $\delta_3=0.010$ in the chain $L_{1111}(2,0,2,0,2,0,2)$ considered in Example  of Sec.\ref{Section:8n} ($L_2(2)\equiv L_1(2)$ in this case) and  $\delta_1=0.247$, $\delta_2=0.977$, $\delta_3=0.018$ in the
  chain $\hat L_{1221}(2,0,2,0,2,0,2)$ considered in   Example 2 of Sec.\ref{Section:def}. 
  }
  \label{Fig:8n_ex1}
\end{figure*}

{\bf Example: the HPSTs in the chain $L_{1111}(2,0,2,0,2,0,2)$.}
We consider the chain of eight nodes as one obtained  by 
joining of two 4-nodes chains $L_{11}(2,0,2)$ constructed in Example 1 of Sec.\ref{Section:4n}, see  Fig.\ref{Fig:8n_ex1} where 
$\delta_2=1$. 

To provide the HPSTs among all nodes we must take a small coupling constant $D_4$ between these two chains. Namely, it must be less than  $\delta=0.196$  introduced in  Example 1 of Sec.\ref{Section:4n}. We take $D_1=D_7=1$, $D_2=D_6=\delta_1=0.196$, $D_3=D_5=\delta_2=1$, $D_4=\delta_3\ll 0.196$.  The disadvantages of this chain are  big parameter $T^{(8)}_{8}$ and comparatively small values of $P^{(8)}_{ij}$, $i,j=1,\dots,8$ ($i\neq j$). 
After optimization we  obtain $\delta_3=0.010$. The parameters of the HPST(8;1,2,3,4,5,6,7,8) are represented in Table 4.

\section{Deformations of the chains improving the parameter $T^{(N)}_{{\cal{N}}}$} 
\label{Section:def}
 It may be shown that the parameters of the HPSTs may be improved varying  all coupling constants in the chains $L_{11}(N_1,M_1,N_1)$ and
$L_{1221}(N_1,M_1,N_1,M_2,N_1,M_1,N_1)$ in a proper way, i.e. varying not only the coupling constants between chains $L_1(N_1)$ (like it was done in Secs.\ref{Section:ch1} and  \ref{Section:ch3}) but also the coupling constants inside of $L_1(N_1)$. This
 allows one to  decrease significantly parameter $T^{(N)}_{{\cal{N}}}$. After such process the above chains
  will be reduced to the
 "optimized" chains which we call $\hat L_{11}(N_1,M_1,N_1)$ and $\hat
 L_{1221}(N_1,M_1,N_1,M_2,N_1,M_1,N_1)$ respectively
(compare  the parameters from Tables 2 and 5 and from  Tables 4 and 6).

 {\bf Example 1: the HPSTs in the chain $\hat L_{11}(3,0,3)$, $P_0=0.9$.}
The purpose of this section is to decrease the parameter $T^{(6)}_{4}$  which have been found in   Example 2 of Sec.\ref{Section:4n}, see Table 2 and Fig.\ref{Fig:d3_4n_ex2}. We take $D_1=D_5=1$, $D_2=D_4=\delta_1$, $D_3=\delta_2$  and vary $\delta_i$, $i=1,2$  keeping in mind the conditions (\ref{condition1},\ref{condition2}).
 We have found that the optimal parameters of the HPST(6;1,3,4,6)  correspond to the $\delta_1=0.769$ and $\delta_2=0.092$, see Table 5 and Fig.\ref{Fig:d3_4n_ex2}.
 
{\bf Example 2: the HPSTs in the chain $\hat L_{1221}(2,0,2,0,2,0,2)$, $P_0=0.8$.}
The parameter $T^{(8)}_{8}$  obtained in Example 1 of Sec.\ref{Section:8n} may be decreased varying the coupling constants in  the chain, see Fig.\ref{Fig:8n_ex1}.
Thus, the optimal parameters of the  HPST(8;1,2,3,4,5,6,7,8) have been found for $\delta_1=0.247$,
$\delta_2=0.977$, $\delta_3=0.018$, see  Table 6 and Fig.\ref{Fig:8n_ex1}.

\begin{table*}[!htb]
\begin{tabular}{|p{1.cm}|p{1.2cm}|p{1.2cm}|p{1.2cm}|p{1.2cm}|p{1.2cm}|p{1.2cm}|p{1.4cm}|}
\hline
 $\!p_i${\LARGE {$\!\!\diagdown$}} $\!\!p_j$
&1         &   2         &  3           &  4  \\\hline
1&            &0.931 3.040 1.196  & 0.904 55.533 -3.010 & 0.912 58.585 -1.806\\\hline
2&0.931 3.040 1.196  &             & 0.906 52.548 1.998 & 0.904 55.533 -3.010\\\hline
3&0.904 55.533 -3.010& 0.906 52.548 1.998&              & 0.931 3.040 1.196 \\\hline
4&0.912 58.585 -1.806& 0.904 55.533 -3.010&0.931 3.040 1.196&
\\\hline
\end{tabular}
\label{Table:HPST(4;4)}
\caption{The parameters $\bar P^{(4)}_{ij}$ (the first number in the box), $\bar t^{(4)}_{ij}$ (the second number in the box) and $\bar\varphi^{(4)}_{ij}$ (the third number in the box) of the HPST(4;1,2,3,4) in $L_{11}(2,0,2)$}
\end{table*}

\begin{table}[!htb]
\begin{tabular}{|p{1.cm}|p{1.2cm}|p{1.2cm}|p{1.2cm}|p{1.2cm}|p{1.2cm}|p{1.2cm}|p{1.4cm}|}
\hline
$\!p_i${\LARGE {$\!\!\diagdown$}} $\!\!p_j$&1         &   3         &  4           &  6  \\\hline
1&            &0.978 11.595 3.120 & 0.909 426.354 0.313& 0.927 414.760 -2.812
\\\hline
3&0.978 11.595 3.120 &             & 0.919 414.762 -2.812& 0.909 426.354 0.313
\\\hline
4&0.909 426.354 0.313 &0.919 414.762 -2.812&              & 0.978 11.595 3.120 \\\hline
6&0.927 414.760 -2.812& 0.909 426.354 0.313  &0.978 11.595 3.120&
\\\hline
\end{tabular}
\label{Table:HPST(6;1346hom)}
\caption{The parameters $\bar P^{(6)}_{p_ip_j}$ (the first number in the box), $\bar t^{(6)}_{p_ip_j}$ (the second number in the box) and $\bar\varphi^{(6)}_{p_ip_j}$ (the third number in the box) of the HPST(6;1,3,4,6) in $L_{11}(3,0,3)$   }
\end{table}

\begin{table}[!htb]
\begin{tabular}{|p{1.cm}|p{1.2cm}|p{1.2cm}|p{1.2cm}|p{1.2cm}|p{1.2cm}|p{1.2cm}|p{1.4cm}|}
\hline
 $\!p_i${\LARGE {$\!\!\diagdown$}} $\!\!p_j$&1         &   2         &  5           &  6  \\\hline
1&            &0.913 3.005 1.147  & 0.926 67.364 2.120 & 0.971 70.375 -3.003\\\hline
2&0.913 3.005 1.147 &             & 0.934 64.400 0.925 & 0.926 67.364 2.120
\\\hline
5&0.926 67.364 2.120 &0.934 64.400 0.925&              & 0.913 3.005 1.147 \\\hline
6&0.971 70.375 -3.003& 0.926 67.364 2.120 &0.913 3.005 1.147 &
\\\hline
\end{tabular}
\label{Table:HPST(6;1256nonhom)}
\caption{The parameters $\bar P^{(6)}_{p_ip_j}$ ((the first number in the box), $\bar t^{(6)}_{p_ip_j}$ (the second number in the box) and $\bar\varphi^{(6)}_{p_ip_j}$ (the third number in the box) of the HPST(6;1,2,5,6) in $L^{C}_{12}(2,2,2)$ }
\end{table}

\begin{table}[!htb]
\begin{tabular}{|p{1.cm}|p{1.4cm}|p{1.4cm}|p{1.4cm}|p{1.4cm}|p{1.4cm}|p{1.4cm}|p{1.4cm}|p{1.4cm}|}
\hline
$\!p_i${\LARGE {$\!\!\diagdown$}} $\!\!p_j$&1         &   2       &  3         &  4         &5             & 6            &  7           &  8
\\\hline
1&              &0.930 3.040 1.181&0.882 55.557 2.666&0.854 52.463 1.573&0.813 3385.361 -2.762&0.835 3382.303 2.367&0.818 3329.745 0.947&0.862 3326.706 -0.245
\\\hline
2& 0.930 3.040 1.181  &           &0.871 52.563 1.431&0.855 55.556 2.666&0.812 3382.302 2.367&0.829 3379.311 1.141&0.886 3326.702 -0.227&0.818 3329.745 0.947
\\\hline
3&0.882 55.557 2.666  &0.871 52.563 1.431&           &0.933 3.044 1.143 &0.830 3329.851 0.754&0.893 3326.800 -0.376&0.829 3379.311 1.141&0.835 3382.303 2.367
\\\hline
4&0.854 52.463 1.573  &0.855 55.556 2.666&0.933 3.044 1.143&            &0.876 3326.807 -0.396&0.830 3329.851 0.754&0.812 3382.302 2.367&0.813 3385.361 -2.762
\\\hline
5&0.813 3385.361 -2.762&0.812 3382.302 2.367&0.830 3329.851 0.754&0.876 3326.807 -0.396&       &0.933 3.044 1.143   &0.855 55.556 2.666  &0.854 52.463 1.573
\\\hline
6&0.835 3382.303 2.367&0.829 3379.311 1.141&0.893 3326.800 -0.376&0.830 3329.851 0.754&0.933 3.044 1.143&          &0.871 52.563 1.431&0.882 55.557 2.666
\\\hline
7&0.818 3329.745 0.947&0.886 3326.702 -0.227&0.829 3379.311 1.141&0.812 3382.302 2.367&0.855 55.556 2.666  &0.871 52.563 1.431&       &0.930 3.040 1.181
\\\hline
8&0.862 3326.706 -0.245&0.818 3329.745 0.947&0.835 3382.303 2.367&0.813 3385.361 -2.762&0.854 52.463 1.573&0.882 55.557 2.666&0.930 3.040 1.181&
\\\hline
\end{tabular}
\label{Table:HPST(8;8nonhom)}
\caption{The parameters $\bar P^{(8)}_{p_ip_j}$ (the first number in the box),
$\bar t^{(8)}_{p_ip_j}$ (the second number in the box) and $\bar\varphi^{(8)}_{p_ip_j}$ (the third number in the box) of the HPST(8;1,2,3,4,5,6,7,8) in $ L^C_{1111}(2,0,2,0,2,0,2)$ }
\end{table}

\begin{table}[!htb]
\begin{tabular}{|p{1.cm}|p{1.2cm}|p{1.2cm}|p{1.2cm}|p{1.2cm}|p{1.2cm}|p{1.2cm}|p{1.2cm}|}
\hline
 $\!p_i${\LARGE {$\!\!\diagdown$}} $\!\!p_j$&1         &   3         &  4           &  6  \\\hline
1&            &0.908 13.095 2.588 & 0.919 87.366 1.483 & 0.916 74.231 -1.044
\\\hline
3&0.908 13.095 2.588 &             & 0.905 74.324 -1.178 & 0.919 87.366 1.483
\\\hline
4&0.919 87.366 1.483 &0.905 74.324 -1.178 &              &0.908 13.095 2.588 \\\hline
6& 0.916 74.231 -1.044 & 0.919 87.366 1.483  &0.908 13.095 2.588&
\\\hline
\end{tabular}
\label{Table:HPST(6;1346nonhom)}
\caption{The parameters $\bar P^{(6)}_{p_ip_j}$ (the first number in the box), $\bar t^{(6)}_{p_ip_j}$ (the second number in the box) and  $\bar\varphi^{(6)}_{p_ip_j}$ (the third number in the box) of the HPST(6;1,3,4,6) in $\hat L_{11}(3,0,3)$  }
\end{table}

\begin{table}[!htb]
\begin{tabular}{|p{1.cm}|p{1.4cm}|p{1.4cm}|p{1.4cm}|p{1.4cm}|p{1.4cm}|p{1.4cm}|p{1.4cm}|p{1.4cm}|}
\hline
 $\!p_i${\LARGE {$\!\!\diagdown$}} $\!\!p_j$&1         &   2       &  3         &  4         &5             & 6            &  7           &  8
\\\hline
1&              &0.893 2.984 1.139&0.858 43.271 -3.070&0.883 46.305 -1.954&0.847 1537.838 -1.667&0.862 1534.792 -2.751&0.812 1488.470 -0.718&0.870 1491.478 0.383
\\\hline
2&0.893 2.984 1.139&              &0.873 46.284 -1.976&0.820 43.267 -3.067&0.823 1534.790 -2.749&0.866 1537.766 -1.600&0.906 1491.433 0.464&0.812 1488.470 -0.718
\\\hline
3&0.858 43.271 -3.070&0.873 46.284 -1.976&             &0.895 3.059 1.064 &0.814 1488.517 -0.836&0.901 1491.551 0.276&0.866 1537.766 -1.600&0.862 1534.792 -2.751
\\\hline
4&0.883 46.305 -1.954&0.820 43.267 -3.067&0.895 3.059 1.064 &            &0.871 1491.604 0.187&0.814 1488.517 -0.836&0.823 1534.790 -2.749&0.847 1537.838 -1.667
\\\hline
5&0.847 1537.838 -1.667&0.823 1534.790 -2.749&
0.814 1488.517 -0.836&0.871 1491.604 0.187&       &0.895 3.059 1.064   &0.820 43.267 -3.067    &0.883 46.305 -1.954
\\\hline
6&0.862 1534.792 -2.751&0.866 1537.766 -1.600&0.901 1491.551 0.276&0.814 1488.517 -0.836    &0.895 3.059 1.064&         &0.873 46.284 -1.976    &0.858 43.271 -3.070
\\\hline
7&0.812 1488.470 -0.718&0.906 1491.433 0.464&  0.866 1537.766 -1.600 &0.823 1534.790 -2.749&0.820 43.267 -3.067&0.873 46.284 -1.976&            &0.893 2.984 1.139
\\\hline
8&0.870 1491.478 0.383&0.812 1488.470 -0.718&0.862 1534.792 -2.751   &0.847 1537.838 -1.667&0.883 46.305 -1.954&0.858 43.271 -3.070&0.893 2.984 1.139&
\\\hline
\end{tabular}
\label{Table:HPST(8;8hom)}
\caption{The parameters $\bar P^{(8)}_{p_ip_j}$ (the first number in the box),
$\bar t^{(8)}_{p_ip_j}$ (the second number in the box) and $\bar\varphi^{(8)}_{p_ip_j}$ (the third number in the box) of the HPST(8;1,2,3,4,5,6,7,8) in $ \hat L^C_{1221}(2,0,2,0,2,0,2)$}
\end{table}

\section{Conclusions}
\label{Section:Conclusions}
Using the numerical simulations we demonstrate that the $N$-nodes  spin 1/2 chains with wide spread of the properly adjusted coupling constants allow the HPSTs between different nodes. It is important that the whole Hamiltonian (\ref{Hamiltonian}) rather then approximation by the nearest neighbour interaction must be used for the correct description of the HPSTs in such chains. 
  
We have found that two spin 1/2 chains $L_1(N_1)$ with the HPST between end nodes may be connected by a relatively weak bond to get a chain with the HPSTs among four nodes, see Sec.\ref{Section:4n}. In turn, having two such chains we may connect them by another weak bond to get a chain with the HPSTs among eight nodes (see Sec.\ref{Section:8n}), and so on. Formally, the number  of the nodes allowing the HPSTs among all of  them  may be 
$2^s$, where $s=1,2,\dots$. 
However, the disadvantage of such chains is a rapid increase of the time interval $T^{(N)}_{{\cal{N}}}$  with the number of nodes involved in the HPSTs. The mechanism decreasing $T^{(N)}_{{\cal{N}}}$ would be 
important for the implementation of these chains.

We also demonstrate that the speedup of the state transfer between the nodes $p_i$ and $p_{i+1}$   separated by the distance $R$ may be achieved using the intermediate chain with properly adjusted coupling constants, see Sec.\ref{Section:4n_mod}.

This work is supported by Russian Foundation for Basic Research through the grant 07-07-00048 and by the Program of the Department of Chemistry and Material Science of RAS No.18.

\end{document}